\documentclass[a4paper]{jpconf}
\usepackage{graphicx}
\usepackage{amsmath}
\usepackage[utf8]{inputenc}
\usepackage[symbol]{footmisc}
\usepackage{hyperref}
\usepackage{lineno}
\begin{document}

\title{Design and development of the ALICE common readout unit user logic firmware for the Muon Identifier readout chain }

%\author{D O Thys-dingou, A Raji, E Z Buthelezi, S V F\"{o}rtsch}

\author{D O Thys-dingou$^1$\renewcommand{\thefootnote}, A Raji$^1$, 
E Z Buthelezi$^{2,3}$ and S V F\"{o}rtsch$^2$}

\address{$^1$ Department of Electrical, Electronic and Computer Engineering, Cape Peninsula University of Technology, Bellville South Industrial, Cape Town, 7530, South Africa}
\address{$^2$ Department of SSC lab, iThemba LABS, Faure, Cape Town, 7100, South Africa}
\address{$^3$ School of Physics, University of the Witwatersrand, Johannesburg, 2000, South Africa}

\ead{dieuveil.orcel.thys-dingou@cern.ch}

\begin{abstract}
A Large Ion Collider Experiment (ALICE) at the Large Hadron Collider (LHC) at CERN went through a major upgrade in which some of its subdetectors were replaced with new ones, while others are equipped with new electronics. The aim of the upgrade is to withstand higher collision rates during the third running period (Run 3), which started in 2022. As part of the upgrade, certain subdetectors such as the Muon Trigger, renamed to Muon Identifier, now operate in a continuous, triggerless readout mode, in addition to the previous triggered readout mode. Due to the increased quantity of data, typical methodologies are impossible to employ without massive efforts to expand the processing capacity. Since the new ALICE computing system cannot keep up with the increased data flow of the Muon Identifier, a new processing algorithm has been established. These proceedings provide a insight to the new approach of processing the Muon Identifier readout data based on a customized user logic FPGA firmware.
\end{abstract}

%\linenumbers

\section{Introduction}
ALICE \cite{ALICE} is one of the four main experiments at the LHC. It is designed to study the strongly interacting matter, namely the quark--gluon plasma (QGP) and its properties. In order to unravel the enigma of the universe, the ALICE detector records data during lead--lead (Pb--Pb), proton--lead (p--Pb), and proton--proton (pp) collisions. Based on data collected during Runs 1 and 2, ALICE is the leading heavy-ion experiment in the world and is quickly expanding the knowledge gathered in previous experiments all over the world. The LHC completed the three-year planned second Long Shutdown, which started at the end of 2018 to prepare for Run 3. In line with the LHC upgrade, the ALICE detector had a major upgrade \cite{LoI}. This upgrade addresses the challenge of reading out Pb-Pb collisions at a rate of 50 kHz, pp, and p-Pb collisions at 200 kHz and higher. At the center of the ALICE upgrade strategy, is a high-speed readout approach based on a Common Readout Unit (CRU), which has been developed for detector data readout, reconstruction, multiplexing, and data decoding on the Online-Offline (O²) computing system.
\newline 
\newline 
Many of the proposed physics observables require a change in the data-taking strategy, moving away from triggering a small subset of events to continuous online processing and recording of all events. To achieve these goals, ALICE has been upgraded in such a way that all interactions will be scrutinized with online precision. The upgrade entailed the replacement of some subdetectors with new ones, making use of new technologies, while others are now equipped with new front-end and readout electronic systems \cite{RD}. Thus far, the selection of single muon and dimuon events with a maximum trigger rate of 1 kHz, limited by readout capabilities, was provided by the Muon Trigger (MTR), as well as muon identification. However, the upgrade strategy described in the Letter of Intent (LoI)\cite{LoI} does not require a muon trigger since all events of interest are now read out upon the interaction trigger before online selections. For this reason, as part of the upgrade, the new Muon IDentifier (MID) subdetector plays the role of muon identifier.

\section{Muon IDentifier}
\begin{figure}[h]
\centering
\includegraphics[width=0.8\textwidth]{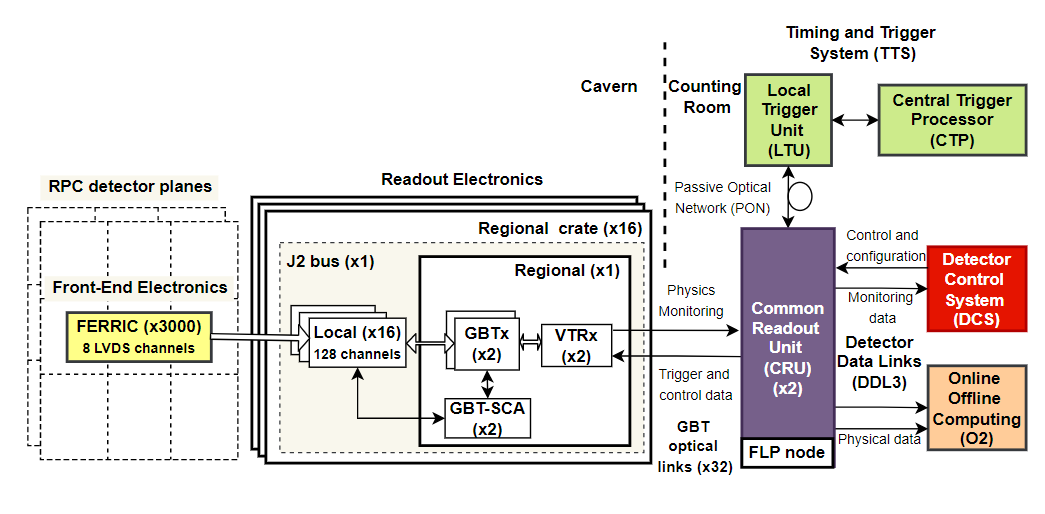}
\caption{A schematic description of the MID readout chain architecture for LHC Run 3.}
\label{fig:mid_ro}
\end{figure}

The MID \cite{MID} is based on 72 single-gap Resistive Plate Chamber (RPC) detectors, arranged in 2 stations of 2 chambers, each at a distance of about 16 m and 17 m from the interaction point, respectively. Its readout chain block diagram is shown in Fig. \ref{fig:mid_ro}.

The MID readout chain consists of about 21,000 strips connected to the 72 RPC detectors spread over multiple Front-End Electronic Rapid Integrated Circuit (FEERIC) cards equipped with one or two customized Application-Specific Integrated Circuits (ASICs). The strip signals from the FEERICs are propagated to the readout electronics using high-speed Low-voltage Differential Signaling (LVDS) channels. The readout electronics (local and regional cards) act as interface between the on and off-detector electronics. They are mounted inside the cavern a little further away from the detector stations, where the radiation is lower. Since the colliding beams will produce a large amount of radiation in the area around the ALICE detector in the cavern, the regional cards are equipped with a radiation hardening Gigabit Transceiver set of chips (GBTx and GBT-SCA) \cite{GBTx, GBT_SCA} used to facilitate the bidirectional connections between the readout electronics and CRUs through optical links, namely, GBT links. The CRUs are the key components of the chain. They combine and multiplex data from multiple readout electronic cards as well as timing and trigger information generated from the Central Trigger Processor (CTP) via the Local Trigger Unit (LTU) before transmitting the data to the O² computing facility for processing and storage. The CRUs are mounted on computers housed in the intermediary computer room, called the counting room, tens of meters above the ALICE cavern and thus do not require to be protected from the radiation, as is the case for the readout electronics. These computers can be reached over the network from the main Detector Control System (DCS). The DCS manages the readout chain by sending commands and monitoring the system. Experimental data are moved from the First Level Processor (FLP) to the Event Processing Node (EPN) for processing and storage. The EPN is an internal component of the O² computing system \cite{O2}.

\section{CRU firmware}
The standard approach of delivering raw data to the O² system is no longer sufficient to meet the needs of the newly enhanced MID subdetector. As a result, an alternative option was presented to the ALICE collaboration \cite{CRU_FW}. Since the release of its first official version in 2018, the CRU firmware provides basic functionalities allowing to interact with multiple systems via a variety of interfaces and read out any subdetector without conducting any first stage data analysis. However, the CRU firmware can be customised to meet the demands of different subdetectors in the ALICE experiment. In particular, a first stage data analysis can be implemented before online and offline reconstruction. This customisation is referred to as "User Logic". It is the responsibility of the MID specialists to decide how data should be handled in their user logic.

\begin{figure}[h!]
\centering
\includegraphics[width= 0.8\textwidth]{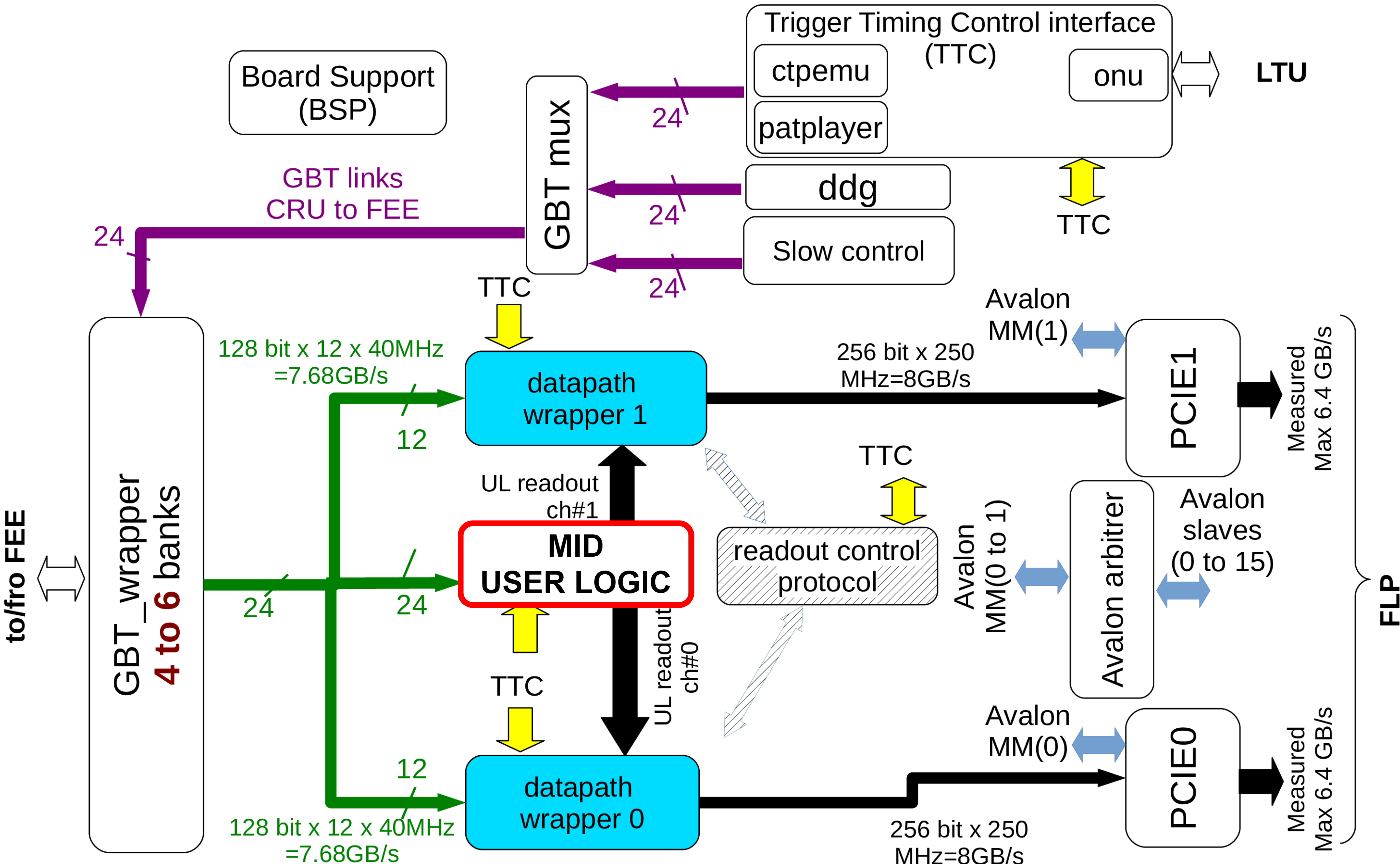}
\caption{CRU firmware architecture. Adapted from \cite{CRU_FW}.}
\label{fig:cru_fw}
\end{figure}

The CRU firmware architecture is illustrated in Fig.\ref{fig:cru_fw}. It is composed of several modules interacting with several interfaces. From left to right, the main interfaces are the GBT wrappers, Board Support Package (BSP), Datapath Wrappers (DWs), Timing and Trigger Control (TTC), Dedicated Data Generator (DDG), slow control, and PCIe endpoints. All of these interfaces provide indispensable functionalities to the CRU firmware, and at the heart of it is a tailored user logic component with functionalities that are unique to MID.

The key general requirements of the user logic component are derived from several articles and are cited in these proceedings, but not duplicated. These requirements are described in \cite{TTC_CRU, CRU_GIT}. One of the main functional requirements of the user logic component is to reduce the data rates transmitted to the O² system by performing zero suppression. It also has to carefully handle the readout electronics anomalies and errors, without stopping the data acquisition of the readout chain. Transient losses of payload (physics) data are not tolerated. This means that at all cost, every level of the user logic hierarchy must be able to identify anomalies from the levels below and continue to transmit packets and follow the communication rules with the levels above. The violation of communication protocols, corruption of data, and unresponsiveness of the system in case of errors shall be prevented.

\section{Architecture and design of the user logic component}

\begin{figure}[h!]
\centering
\includegraphics[width= 0.85 \textwidth]{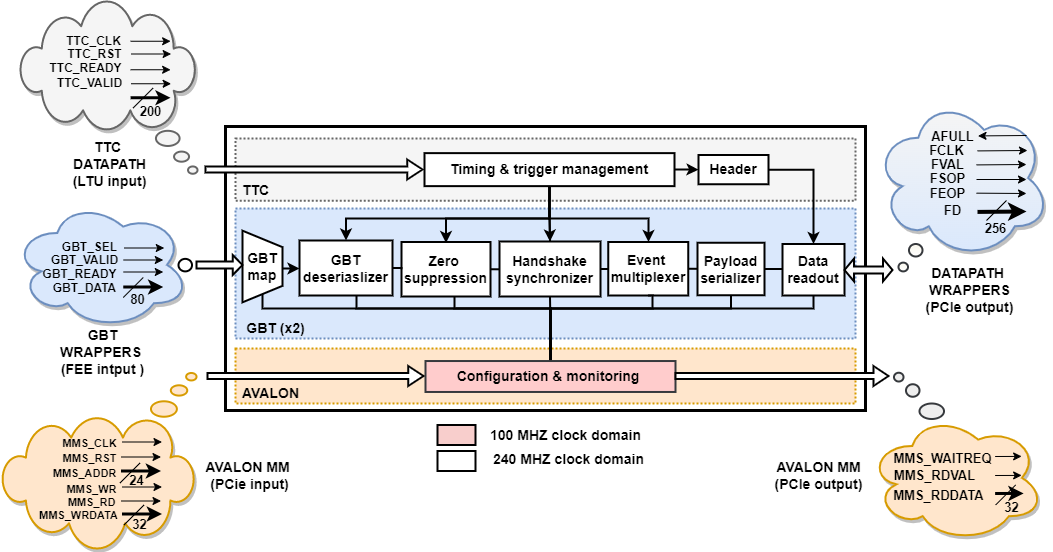}
\caption{Structure of the user logic design showing the three main segments and data flow.}
\label{fig:ul_design}
\end{figure}

The user logic is designed in a sequential manner, with all processing occurring one after another from an input--output perspective. The objective of this approach is to facilitate error tracking. The user logic consists of three main segments, each of which is linked to a specific interface of the CRU firmware (see Fig.\ref{fig:cru_fw}). A representation of the user logic block diagram and its interfaces is shown in Fig.\ref{fig:ul_design}. Starting from the top is the TTC segment (grey), which receives data from the timing and trigger system through the TTC interface. Next is the GBT segment (blue), which receives data from the readout electronics via the GBT wrappers, analyses them then combines them with the Raw Data Header (RDH) extracted from timing and trigger information before transmitting them to the O² system via the datapath wrappers. The GBT segment is the only part of the design that can be duplicated through parameterization. Hence, enabling the possibility to process multiple GBT links, allows for improvement and adaption to diverse testing scenarios. The last segment is the Avalon (orange), which provides configuration and monitoring through the PCIe interface.

\section{Test-bench layout}
\begin{figure}[h!]
\centering
\includegraphics[width= \textwidth]{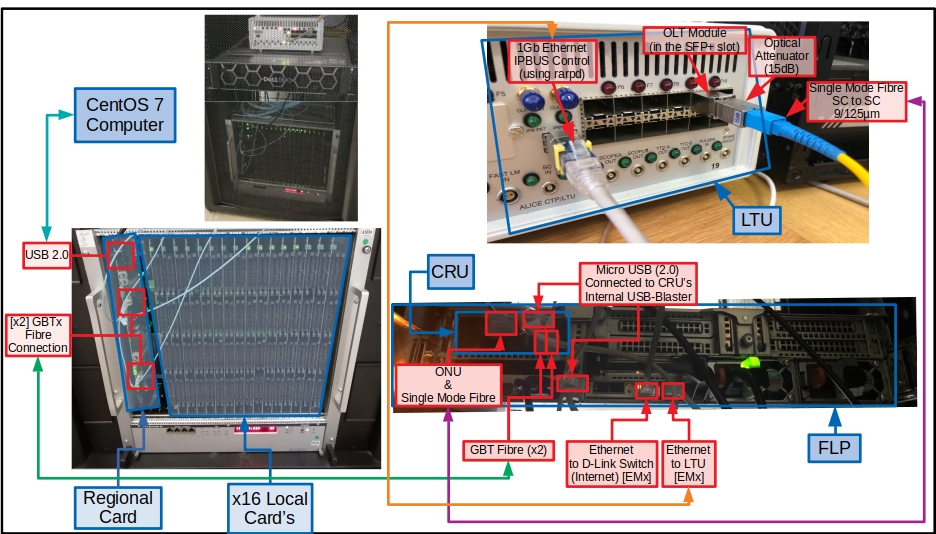}
\caption{New test bench located in the laboratory S64 at iThemba LABS. The diagram shows the readout chain components and how they are connected.}
\label{fig:testbench}
\end{figure}

A test bench at iThemba LABS was implemented to extend the capabilities of the user logic and match the test scenarios performed with readout electronics at SUBATECH, Nantes (where the readout electronics cards were designed \cite{WEB}). The test bench is a scaled-down replica of the MID readout chain without the RPC detectors. It includes a fully-equipped VME crate (16 x local, 1 x regional, and 1 x J2 bus boards), the LTU, CRU, and FLP.  Figure \ref{fig:testbench} depicts the test bench setup and illustrates how various components are linked together. The full setup can be seen on the top left, while in the bottom left the fully-equipped VME crate is shown. The local and regional cards are plugged into the crate via the J2 bus card sitting at the back of the crate. Three cables are exiting the regional card, two of which are optical cables, and connect the regional card to the CRU. The latter is a USB (2.0) cable connected to a CentOS PC, which is used to configure and program the local and regional FPGAs. At the top-right is the LTU, which can be used to interact with multiple CRUs via a splitter. However, for this application only one CRU is needed. The connection between the LTU and the CRU is done via a single-mode SC to SC optical cable. The LTU uses an Ethernet cable to interact with the FLP software, which runs on CentOS 8. Finally, on the bottom-right is the CRU board housed by the FLP server. The CRU board is internally attached to the FLP via the PCIe connectors, its FPGA can be programmed using the PCIe interface or via its integrated USB blaster programmer, which connects to the FLP server using a micro-USB (2.0) cable.

\section{Results}
\begin{figure}[h!]
\centering
\includegraphics[width= 0.99 \textwidth]{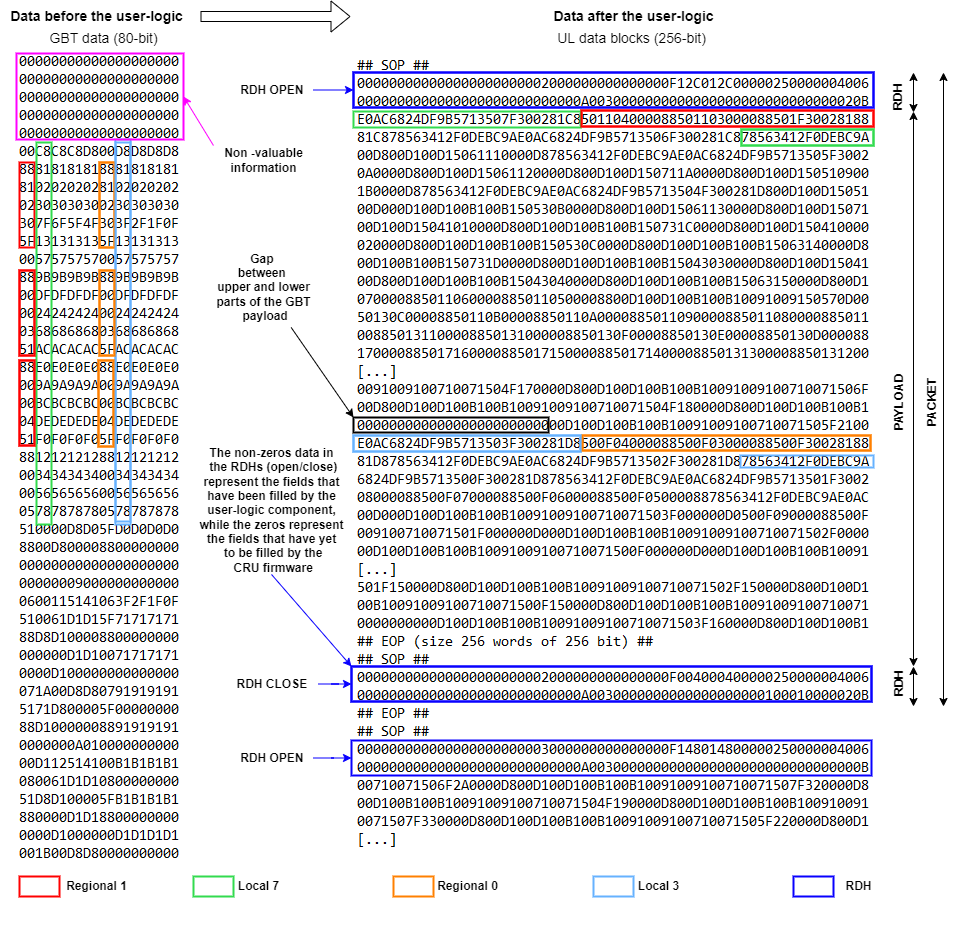}
\caption{Simulation results extracted from ModelSim. This image compares data before and after the user logic processing. As it can be seen, the packets start and end with SOP and EOP signals. Each packet includes RDHs (open and close), that specify the borders of the packet.}
\label{fig:result}
\end{figure}

The simulated results of the working user logic firmware were generated in ModelSim using a simulation framework provided by SUBATECH. A subset of the results is illustrated in Fig.\ref{fig:result}. As can be observed, the user logic output packets fulfill the readout control protocol criteria specified in \cite{CRU_GIT}, which states that packets created by the user logic must begin and terminate with the Start Of Packet (SOP) and End Of Packet (EOP) signals, and each packet must be enclosed by the RDH. The output data format transmitted by the user logic is based on the GBT raw data, pre-analysed and concatenated into multiple data blocks of 256-bit to form the payload included in the packets. \newline

A clear comparison of data before and after being processed by the user logic is illustrated in Figure \ref{fig:result}. It shows the outcome result of the data before and after all processing stages the user logic firmware have been completed. These simulation results validate the aim of this research, which is to enhance the way data are processed by only transmitting valuable information to the FLP. The hardware test results did not reflect the simulation results at first, but were refined after each iteration until complete accuracy was achieved. The user logic helped decreasing the amount of data transmitted from the CRU to the FLP by roughly 80\%, and improved the readability of the data at the O² level.

\section{FPGA resource usage}
The CRU firmware combined with the user logic component use about 160k (38\%) Adaptive Logic Modules (ALMs) and 1355 (50\%) RAM blocks of the available resources. These results were obtained after integrating and compiling the CRU firmware with the user logic component. Table \ref{resource} provides a summary of the total FPGA resource used. The user logic consumes around 37k (9\%) ALMs and 271 (10\%) RAM blocks of the overall resources. These findings meet the requirement of this study, but are not good enough as the long-term aim is to process data from 16 GBT links while maintaining the overall RAM consumption below 75\%.

%\begin{center}
\begin{table}[h]
\caption{\label{resource} FPGA resource usage of the CRU firmware after insertion of the user logic component.}
%\footnotesize\rm
\centering
\begin{tabular}{@{}*{7}{l}}
\br
Resource name&Total in ratio& Total in percentage\\
\mr
\verb"Logic utilization (in ALMs)"&160,282 / 427,200&38\% \\
\verb"Pin"&369 / 960&38\% \\
\verb"Block memory bits"&19,982,660 / 55,562,240&36\% \\
\verb"RAM blocks"&1,355 / 2,713&50\% \\
\verb"RX channels"&41 / 72&57\% \\
\verb"TX channels"&41 / 72&57\% \\
\verb"Phase Locked Loops (PLLs)"&59 / 144&41\% \\
\br
\end{tabular}
\end{table}
%\end{center}

\section{Conclusion}
The user logic firmware performed reasonably well during the simulation and hardware tests. The findings shown in these proceedings demonstrate that the user logic is stable, reliable, and built to read out Pb--Pb collisions at a rate of 50 kHz, and pp and p--Pb collisions at 200 kHz and higher without any issues. The methodology implemented shows that it is feasible to considerably reduce the data bandwidth transmitted from the CRU to the FLP by roughly 80\%. The user logic has passed the validation tests and so fulfils the MID requirements. The results obtained also indicate that with some optimizations, the user logic can immensely contribute to the development of a full scale user logic component capable of processing data from the complete MID system during the LHC Run 3.


\section*{References}
\begin{thebibliography}{9}
\bibitem{ALICE} ALICE Collaboration 2008, The ALICE experiment at the CERN LHC, {\it Journal of Instrumentation}, {\bf 3}: S08002."
\verb"DOI:10.1088/1748-0221/3/08/S08002"

\bibitem{LoI} ALICE Collaboration 2014, Upgrade of the ALICE Experiment: Letter of Intent, {\it Journal of Physics G: Nuclear and Particle Physics} {\bf 41}. \verb"DOI:10.1088/0954-3899/41/8/087001"
 
\bibitem{RD} 
Antonioli P., Kluge A. and Riegler W. 2013, Upgrade of the ALICE Readout and Trigger System, CERN-LHCC-2013-019, ALICE-TDR-015. 

\bibitem{MID} Terlizzi L., The ALICE Muon IDentifier (MID), {\it Journal of Instrumentation}, 2020, {\bf 15} C10031.
\verb"DOI:10.1088/1748-0221/15/10/C10031"

\bibitem{GBTx}  Moreira P. et al. 2009, The GBT Project, {\it Topical Workshop on Electronics for Particle Physics}.
\verb"DOI:10.5170/CERN-2009-006.342"

\bibitem{GBT_SCA} Caratelli A. et al. 2015, The GBT-SCA: A radiation tolerant ASIC for detector control and monitoring applications in HEP experiment, {\it Journal of Instrumentation}, {\bf 10} C03034. 
\verb"DOI:10.1088/1748-0221/10/03/C03034"

\bibitem{O2} 
Buncic P., Krzewicki M. and Vande Vyvre P. 2015, Technical Design Report for the Upgrade of the Online-Offline Computing System, CERN-LHCC-2015-006, ALICE-TDR-019.

\bibitem{CRU_FW} 
Bourrion O., Bouvier J. et al. 2021, Versatile firmware for the Common Readout Unit
(CRU) of the ALICE experiment at the LHC, {\it Journal of Instrumentation}, {\bf 15} P05019. 
\verb"DOI:10.1088/1748-0221/16/05/P05019" 

\bibitem{TTC_CRU} 
Kvapil J. et al. 2021, ALICE Central Trigger System for LHC Run 3,
 {\it Conference on High-Energy Physics}, {\bf 251} 04022. 
\verb"DOI:10.1051/epjconf/202125104022" 

\bibitem{CRU_GIT} 
Bourrion O. et al., CRU firmware private GitLab project. \verb"https://gitlab.cern.ch/alice-cru/cru-fw"

\bibitem{WEB} 
Renard C., MID Readout electronics project.\newline
\verb"http://www-subatech.in2p3.fr/~electro/projets/alice/dimuon/trigger/upgrade/index.html"

\end{thebibliography}
\end{document}